\documentclass[aps,prd, preprintnumbers,groupedaddress,nofootinbib]{revtex4}
\usepackage{graphicx}
\usepackage{latexsym}
\usepackage{amsfonts}
\usepackage{amssymb}
\usepackage{amsmath}
\usepackage{slashed}
\usepackage{feynmp}
\usepackage{hyperref}
\usepackage{xspace}

\newcommand{\tev}{\ensuremath{\mathrm{\,Te\kern -0.1em V}}\xspace}
\newcommand{\gev}{\ensuremath{\mathrm{\,Ge\kern -0.1em V}}\xspace}
\newcommand{\tevt}{\ensuremath{\mathrm{Te\kern -0.1em V}}\xspace}
\newcommand{\kev}{\ensuremath{\mathrm{\,ke\kern -0.1em V}}\xspace}
\newcommand{\mev}{\ensuremath{\mathrm{\,Me\kern -0.1em V}}\xspace}

\begin{document}

\title{Xogenesis}

\author{Matthew R.~Buckley,$^{1,2}$  and Lisa Randall$^3$}
\affiliation{$^1$Center for Particle Astrophysics, Fermi National Accelerator Laboratory, Batavia, IL 60510}
\affiliation{$^2$Department of Physics, California Institute of Technology, Pasadena, CA 91125, USA}
\affiliation{$^3$Harvard University, Cambridge, MA 02138, USA}
\preprint{CALT 68-2782,FERMILAB-PUB-10-349-T}
\date{\today}

\begin{abstract}
We present a new paradigm for dark matter in which a dark matter asymmetry is established in the early universe that is then transferred to ordinary matter. We show this scenario can fit naturally into weak scale physics models, with a dark matter candidate mass of this order. We present several natural suppression mechanisms, including bleeding dark matter number density into lepton number, which occur naturally in models with lepton-violating operators transferring the asymmetry that reduce the number density and allow for dark matter much heavier than baryon masses.
\end{abstract}

%\pacs{}

%%%%%%%%%%%%%%%%%%%%%%%
\maketitle

\section{Introduction \label{sec:intro}}

After years of study, the nature of dark matter remains a mystery. While we hope that data will soon decide the issue, at the moment we have only theoretical clues and experimental constraints. Thermal freeze-out calculations suggest a connection to the weak energy scale as stable weak scale particles have approximately the right abundance to be viable candidates for dark matter.  However, we don't yet know whether the weakly interacting, massive particle (WIMP) hypothesis is correct. In particular, detailed calculations point to tunings that are essential in explicit WIMP examples in order to get a thermal abundance in agreement with observations, while recent theoretical work has presented many new non-canonical dark matter candidates that offer viable alternatives to the WIMP paradigm.

Given the uncertainties in the nature of dark matter, it is worth noting another remarkable coincidence: the closeness of the dark matter energy density to that of ordinary matter, differing by only a factor of about six.  This remarkable fact is suggestive of an underlying connection between the origin of both baryons and dark matter.

This relationship has been exploited in models of Asymmetric Dark Matter (ADM) (see Refs.~\cite{Kaplan:1991ah,Thomas:1995ze,Hooper:2004dc,Agashe:2004bm,Kitano:2004sv,Cosme:2005sb,Farrar:2005zd,Suematsu:2005kp,Tytgat:2006wy,Banks:2006xr,Kitano:2008tk} and more recently, in Refs.~\cite{Kribs:2009fy,An:2009vq,Cohen:2009fz,Kaplan:2009ag,Cohen:2010kn}), 
in which dark matter has essentially the same number density as ordinary matter. In accordance with the measured energy densities, this requires the dark matter to be light -- approximately $5-10\gev$. In most of these models, it is assumed that a baryon asymmetry is created and somehow transferred to dark matter, although very recent papers suggest the opposite \cite{Shelton:2010ta,Davoudiasl:2010am,Haba:2010bm}.

In this paper we present an alternative framework in a  very general scenario that we call ``Xogenesis.'' 
In such models, a dark matter asymmetry (consisting of a particle $X$) is created that is transferred through dark matter-ordinary matter interactions to the Standard Model (SM) sector. In this paper, we don't present explicit mechanisms for dark matter asymmetry creation, but simply note that many mechanisms of baryogenesis have obvious generalizations to a non-Standard Model sector, presumably with fewer constraints given dark matter's inaccessibility. We leave explicit realizations of this aspect to future work.

\begin{figure}[ht]
\includegraphics[width=0.55\columnwidth]{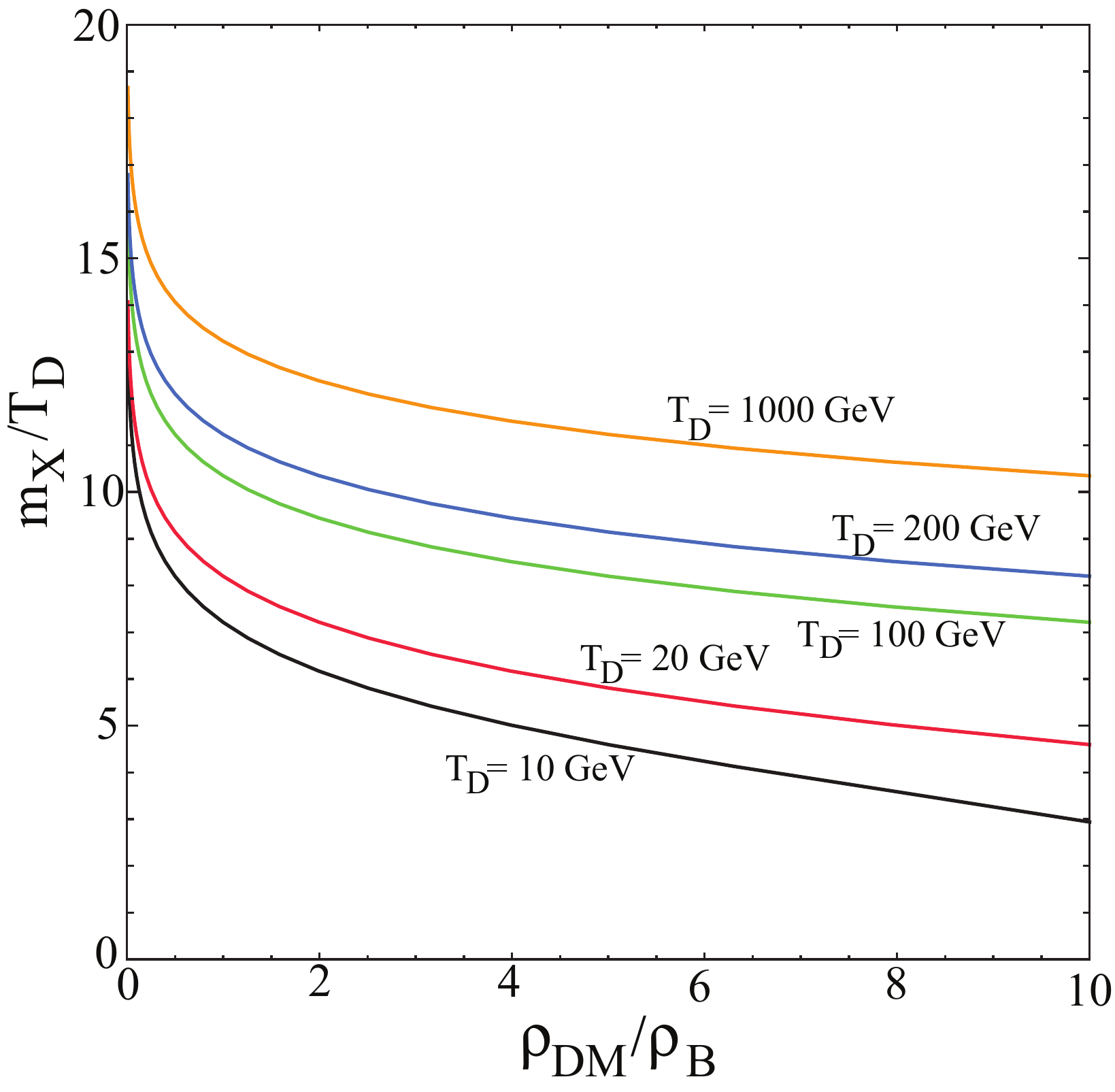}
\caption{The ratio of dark matter energy density $\rho_{\rm DM}$ to baryon energy density $\rho_{B}$ as a function of dark matter mass $m_X$ in units of the temperature at which the $B-X$ transfer decouples $T_D$, for labeled values of $T_D$. As light solution (corresponding to $m_X/T_D \sim 0$ is not shown. See Section~\ref{sec:models} and Eq.~(\ref{eq:su2lsolution}) for detailed explanation. The observed ratio of $\rho_{\rm DM}/\rho_B$ is $5.86$ \cite{pdg}.   \label{fig:massvsrho}}
\end{figure}

A key difference in our approach compared to Refs.~\cite{Shelton:2010ta,Davoudiasl:2010am,Haba:2010bm} is that we assume the dark matter candidate particle is heavy, with mass at or around the weak scale. The biggest potential objection to such dark matter is that it would seem to require exponential tuning to get the necessary suppression of dark matter number density so that the energy in dark matter is not too high. However, this is not the case. Thermal suppression turns on relatively slowly, with exponential behavior becoming apparent only when mass scales differ by more than an order of magnitude. When mass scales are within a natural range -- differing by an order of magnitude or so -- the correct dark matter density is readily achievable. Of course, getting the dark matter density precisely right requires a specific relation among parameters. But this is not a violation of naturalness \cite{Anderson:1994dz}, but merely a fitting of parameters.   For any value in this range, one would find reasonably equivalent energy densities of dark matter and ordinary matter.
This is illustrated in Figure~\ref{fig:massvsrho}, where we show the relative density of dark and ordinary matter as a function of the ratio of the dark matter mass to the decoupling temperature. We see a broad regime where the function is approximately linear. We also note that this mechanism seems to favor lower decoupling temperature for the lightest and most natural dark matter candidates.

As stated, in this paper we simply assume early Universe processes create a dark matter asymmetry. The models we present contain mechanisms for transferring the asymmetry to ordinary matter and we study the constraints this imposes. We have considered several possible classes of transfer mechanisms. Broadly speaking, we divide these into models in which dark matter is charged under $SU(2)_L$ and transfers the asymmetry to ordinary matter via sphalerons, models in which both ordinary matter and dark matter are charged under some new gauge group and exotic sphalerons transfer the asymmetry, models in which $B$-violating operators transfer the asymmetry, and models in which $L$-violating operators do the job. 

These mechanisms have been successfully applied to generate the relevant energy densities in the context of an existing baryon asymmetry being transferred to light dark matter, though mechanisms named darkogenesis \cite{Shelton:2010ta} and hylogenesis \cite{Davoudiasl:2010am} have also been suggested which transfer the asymmetry in the opposite direction. If, on the other hand, dark matter is not relativistic at the temperature $T_D$ at which the $X$-transfer operators decouple, then the number density of dark matter is suppressed. In general, we find when the ratio $m_X/T_D$ is about 10, we get the required density of dark matter compared to baryons in the Universe. This thermal suppression is a generic feature, allowing heavy dark matter in many scenarios of Xogenesis.  

We also discuss two other reasons that dark matter number density might be suppressed relative to baryon number so that dark matter can naturally be weak scale in mass.. In the first, the $SU(2)_L$ sphaleron transfer is only active for a bounded temperature range between the masses of two doublets whose net number density would cancel if they were degenerate \cite{graham}. In the second,  excess $X$-number is bled off into leptons. That is, even after the baryon asymmetry is established (possibly at the sphaleron temperature where a lepton asymmetry gets transferred into an asymmetry in the baryon sector), $X$- and lepton-number violating operators are still in thermal equilibrium allowing $X$ number density to be reduced while lepton number density is increased.     Both these mechanisms cause the transfer to baryons to not be active for the entire temperature range down to $T_D$ when the $X$-number violating operators decouple.

Xogenesis models must also remove the symmetric thermally produced dark matter component, so that the asymmetric component dominates.  When the transfer mechanism is due to higher order operators, the operators necessary to transfer the asymmetry may also lead to the annihilation of this component. In other examples, new interactions are assumed, which in some cases also lead to detectable signatures. A new non-abelian $W'$ with masses much below $m_W$ allows the dark matter to annihilate into dark gauge bosons, but with few --  if any -- direct detection constraints and probably no visible signatures in the near future. Annihilation via a light $Z'$ that mixes with the photon allows the chance for direct detection, depending on the size of the mixing parameter.  While not strictly necessary, the photon-$Z'$ mixing is a generic property, and may be accessible in beam experiments \cite{Bjorken:2009mm}.

We also note one additional constraint that applies to supersymmetric models in which higher dimension operators link $X$ to $L$ or $B$ via the lepton or baryon superpartners. In these cases, the neutralinos that come from the superpartner decay must also be eliminated via self-annihilation. This generally implies that the neutralino should be primarily wino so that  the annihilation cross section is sufficiently large to make   the neutralino component of dark matter a small percentage of the total.  

\section{Transfer Mechanisms \label{sec:models}}

The scenario we propose is quite general.  We illustrate the idea with four classes of baryon-dark matter transfer scenarios: $SU(2)_L$ sphalerons, sphalerons of a new gauge group, and higher-dimensional operators that violate either lepton or baryon number. In each case, chemical equilibrium between dark matter and baryons is maintained until the operators that transfer the dark matter $X$ number into baryon number $B$ decouple at a temperature $T_D$.  Since the net number density $n_i - \bar{n}_i$ of particle species $i$ is proportional to its chemical potential $\mu_i$, the equality of $\mu_X$ and $\mu_B$ would seem to imply $n_X\sim n_B$ (up to ${\cal O}(1)$ numbers that depend on the details of a particular model). Indeed a solution of this sort generally exists when $m_X \sim 5-6 m_{\rm proton}$.  However,   the equation for net number density of particle $i$ (with $g_i$ degrees of freedom) at temperature $T$ and scale factor $R(T)$ tells us the relationship is more subtle
\begin{equation}
n_i = g_i f(m_i/T) T^2 R(T)^3 \mu_i. \label{eq:numberdensity}
\end{equation}
and that a second solution is possible. 

The function $f(x)$ in Eq.~(\ref{eq:numberdensity}):
\begin{equation}
f(x) = \frac{1}{4\pi^2} \int_0^\infty \frac{y^2}{\cosh^2\left(\frac{1}{2}\sqrt{x^2+y^2}\right)}dy \label{eq:f}
\end{equation}
measures the departure from thermal equilibrium of the particle. If the ratio of dark matter mass $m_X$ to $T_D$ is large, then as the transfer operator decouples, the dark matter itself is going out of thermal equilibrium, and has suppressed number density ($f(m_i/T_D) \to 0$ as $m_i/T_D \to \infty$). This results in a lower $n_X$ than would occur if $m_X \ll T_D$, and thus in order to get the right energy density of dark matter a larger dark matter mass is allowed. It should also be noted that the number density is can continue to loosely track the equilibrium number density even after decoupling (that is, the number density to entropy ratio $Y(T_D) \neq Y_\infty$). Therefore, we can expect some additional dilution of the dark matter number density from $T_D$ to the present day. For the purposes of this paper, we ignore this effect.\footnote{The authors thank Yanou Cui and Brian Shuve for this observation}

We see that for a model with a particular value of $T_D$,   two $m_X$ solutions that give the correct dark matter density typically exist: $m_X\sim 5\gev$ (the relativistic solution), and $m_X \sim 10 T_D$ (the non-relativistic solution), though additional scenarios with even lower ratio of $m_X$ to $T_D$ are possible, as we outline below. As the details of a particular model tend to affect only the proportionality constant between $\mu_X$ and $\mu_B$, the non-relativistic solution (dependent as it is on an exponential suppression) is relatively model independent. We now present this calculation.

\subsection{$\mathbf{SU(2)_L}$ Sphalerons}

We begin with perhaps the simplest model, in which the dark matter particle $X_L$ is an $SU(2)_L$ fermionic doublet with hypercharge $+1/2$. The simplest version of this model is ruled out by direct detection constraints, however it is remains useful as a demonstration of the general Xogenesis technique. We shall then consider modifications of the simple $SU(2)_L$ sphaleron model that evade direct detection constraints. 

In this first model, there must be a second fermion charged under $SU(2)_L$ to avoid an $SU(2)$ anomaly \cite{Witten:1982fp}. This second state can be either heavier or lighter (in latter case it must be unstable). Here we will consider only the state relevant to dark matter. In order to give mass to both the charged and neutral states of this chiral fermion, there must be two $SU(2)_L$ singlets as well: $\bar{X}_S^0$ and $\bar{X}_S^-$.   
\begin{equation}
{\cal L} \supseteq y_X X_L \phi \bar{X}^0_S+y_X' X_L \phi^* \bar{X}_S^- + m_0 X_S^0 \bar{X}_S^0. \label{eq:su2lLagrangian}
\end{equation}
As we shall see, this simplest model violates bounds from direct detection (note that adding a $\sim 100\kev$ Majorana mass to allow for inelastic scattering does not work, as such mass terms violate the $X$ symmetry). We will address possible solutions to this issue later, but proceed with the simple model to demonstrate the general calculations in an Xogenesis scenario.

Since the left-handed fermion is charged under $SU(2)_L$, dark matter can be created and destroyed in the $SU(2)_L$ sphaleron. In the SM, the action of the sphaleron creates or destroys baryons and leptons, while preserving the linear combination $B-L$.   Generalizing to $N_X$ families of dark matter (that is, $N_X$ $SU(2)_L$ doublets), we see that the linear combination $B-3/N_X X$ is also preserved. Therefore, the action of the sphaleron enforces chemical equilibrium between $X$ and the quarks, with
\begin{equation}
\mu_{X_L} = - 3N_X \mu_{u_L}. \label{eq:chemeqsu2l}
\end{equation}

Combining Eqs.~(\ref{eq:numberdensity}) for baryons and dark matter with Eq.~(\ref{eq:chemeqsu2l}), and assuming massless quarks, we find that
\begin{eqnarray}
n_X & = & N_X^2 \frac{f(m_X/T_D)}{f(0)} n_B \\
f(m_X/T_D) & = &  \frac{f(0)}{N_X^2} \frac{\rho_{\rm DM}}{\rho_B} \frac{m_{\rm proton}}{m_X}. \label{eq:su2lsolution}
\end{eqnarray} 
Here, $T_D$ is the temperature at which the sphaleron is no longer active. Exact calculation of this value is difficult, so for the purposes of this paper, we assume that it occurs at the Higgs vev $v\sim 200\gev$. 

\begin{figure}[th]
\includegraphics[width=0.6\columnwidth]{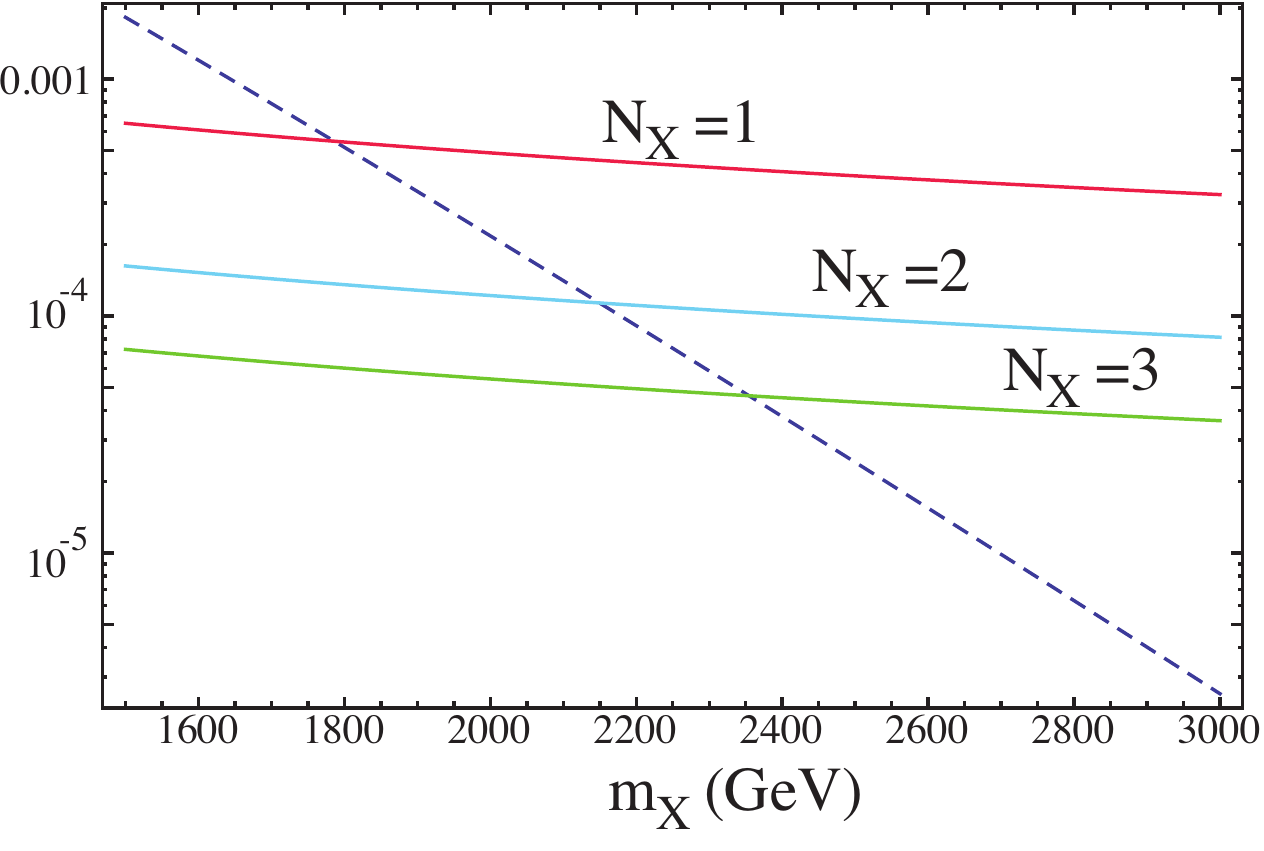}

\caption{Numeric solution to Eq.~(\ref{eq:su2lsolution}, for one, two, or three fermionic dark matter doublets ($N_X=1,2,3$) and assuming a $SU(2)_L$ sphaleron decoupling temperature $T_D = 200\gev$. The blue dotted line is the left-handed side of Eq.~(\ref{eq:su2lsolution}), {\it i.e.}~$f(m_X/T_D)$. \label{fig:su2lsolution}}
\end{figure}

Eq.~(\ref{eq:su2lsolution}) must be solved numerically. Taking the current WMAP values for the energy density of dark matter and baryons, the ratio $\frac{\rho_{\rm DM}}{\rho_B} = 5.86$. In Fig.~\ref{fig:su2lsolution}, we plot the left- and right-hand sides of Eq.~(\ref{eq:su2lsolution}) for $N_X=1,2,3$. The solution for $N_X=1$ is $m_X \sim 1800\gev$, or $9T_D$. As can be seen, the value of $m_X$ which provides the correct dark matter density depends only weakly on the ${\cal O}(1)$ number in the equations for chemical equilibrium (in this case, $3N_X$). 

For the model to work, we require both  a sufficient number of baryons from an initial dark matter asymmetry (while maintaining the observed dark matter density) created by the sphaleron, and a sufficient suppression of the thermal component.  Some process must act to efficiently annihilate the thermal {\it symmetric} dark matter number density. In the current example, it is natural to consider annihilation through $SU(2)_L$ interactions. However the cross section for this process is too small; an $SU(2)_L$ fermion produces the dark matter density when $m_X \sim 1\tev$; larger masses yield too much thermal dark matter \cite{Cirelli:2005uq}.

In order for the sphaleron to change the net $X$ number, the dark matter must be chiral, and so a coupling to the SM Higgs is necessary to provide a mass term, as in Eq.~(\ref{eq:su2lLagrangian}). The large mass required to match observations of $\rho_{\rm DM}$ requires a $y_X \sim 10$ -- near the perturbativity limit. Although perhaps theoretically undesirable, such a large Yukawa yields an efficient annihilation of $X\bar{X}$ pairs into SM fermions. The thermal abundance is given by \cite{Kolb:1990vq}
\begin{equation}
\Omega_{\rm DM} h^2 \approx \frac{1.04 \times 10^9 x_f}{m_{\rm Pl}\sqrt{g_*}(a+3b/x_F)} \label{eq:thermal} 
\end{equation}
where $x_f = m_X/T_f \sim 20$ is the ratio of mass to temperature at freeze-out, $g_*$ is the number of degrees of freedom active at $T_f$, and $a$ and $b$ are, respectively, the $s$- and $p$-wave contributions to the thermally averaged annihilation cross section $\langle \sigma v\rangle$. Roughly speaking, the observed value of $\Omega_{\rm DM}$ occurs when $\langle \sigma v\rangle$ is 1~pb. For $X\bar{X} \to t\bar{t}$, via the SM Higgs, annihilation proceeds only through $p$-wave processes, and
\begin{equation}
b \sim \frac{1}{4\pi} \frac{y_X^2y_t^2 m_X^2}{m_H^4}. \label{eq:higgsb}
\end{equation}
Thus, in order to suppress the symmetric component of dark matter, it must be true that
\begin{equation}
6 \times 10^{-8}\gev^{-2} \ll b \sim 0.3 \gev^{-2} \left(\frac{y_X}{10}\right)^2\left(\frac{y_t}{1}\right)^2 \left(\frac{m_X}{1800\gev} \right)^2\left(\frac{m_H}{100\gev}\right)^{-4}. \label{eq:higgsconstraint}
\end{equation}
Therefore Higgs-mediated annihilation suffices to remove the symmetric component in this scenario.

The required mass as derived in Eq.~(\ref{eq:su2lsolution}) places us in direct conflict with direct detection experiments. The null results from  XENON100 \cite{Aprile:2010um} and CDMS-II \cite{Ahmed:2009zw} rule out neutral component of a $SU(2)_L$ doublet in the $100\gev$--$\tev$ mass range by $\sim 4$ orders of magnitude. The presence of a $\sim 10\kev$ mass splitting induced by a mixing with a Majorana singlet would evade this constraint \cite{Hall:1997ah,TuckerSmith:2004jv}, however such a mechanism would destroy the $X$-number asymmetry and so cannot be present in this model. 

Mixing with Dirac singlet can reduce the direct detection signal by an amount proportional to the fourth power of the doublet component $\epsilon$ of the lightest state in the dark sector. Hence, to avoid conflict with experiment, $\epsilon$ must be $\lesssim 10^{-1}$. However, a dark sector consisting of one left-handed doublet $X_L$ and a Dirac singlet $X_1$ will have two mass eigenvalues, the lightest of which has $\epsilon \gtrsim \tfrac{1}{2}$.

We therefore consider the simplest dark sector that captures the necessary phenomenology: a left-handed doublet $X_L$, and three Dirac gauge singlets: one left-handed ($X_1$) and two right-handed ($X_2$ and $X_3$). After EWSB, the most general Dirac mass matrix for the neutral states is given by the Lagrangian
\begin{equation}
{\cal L} \supseteq y_1 v X_L \bar{X}_2 + y_2 v X_L \bar{X}_3 + m_{12} X_1\bar{X}_2 + m_{13} X_1\bar{X}_3 + \mbox{h.c.}
\end{equation}
Of course, arbitrary choices of the parameters $y_1$, $y_2$, $m_{12}$, and $m_{13}$ will not provide three non-zero mass eigenstates with a lightest state that has a sufficiently small doublet component. 

One possible choice of parameters is as follows: define $y_1 v \equiv M$ (this will become the mass scale of the primarily doublet states), the small parameter $\epsilon$ is defined as $y_1/y_2$, and we choose $m_{13 }\sim 0$ and $m_{12} \sim \epsilon M$. Then the lightest state is a linear combination of $\sim \epsilon X_L + \epsilon X_1 + X_2$ and has a mass $\sim \epsilon^2 M$. 

\begin{figure}[th]
\includegraphics[width=0.39\columnwidth]{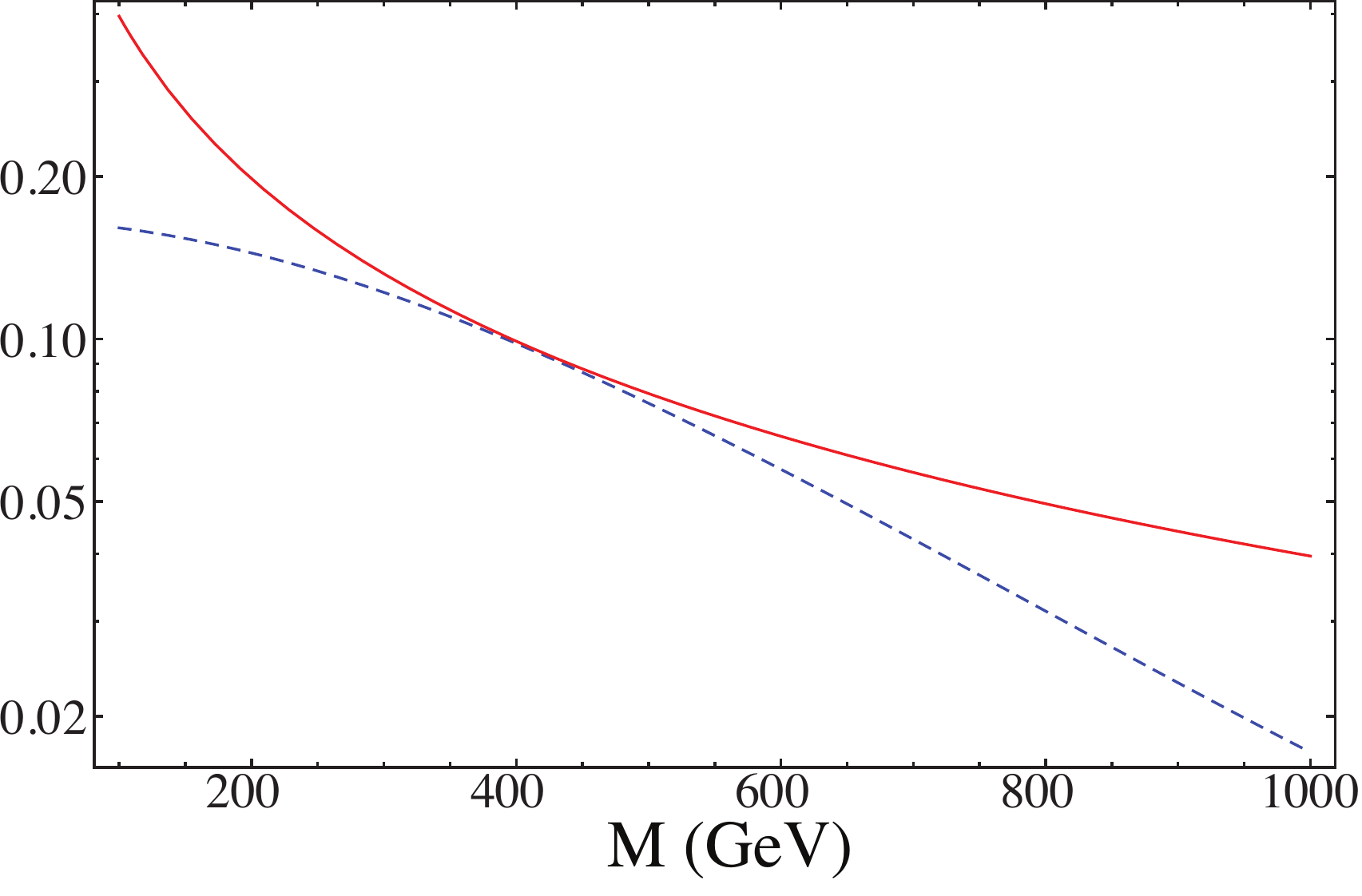}\hspace{2em}\includegraphics[width=0.4\columnwidth]{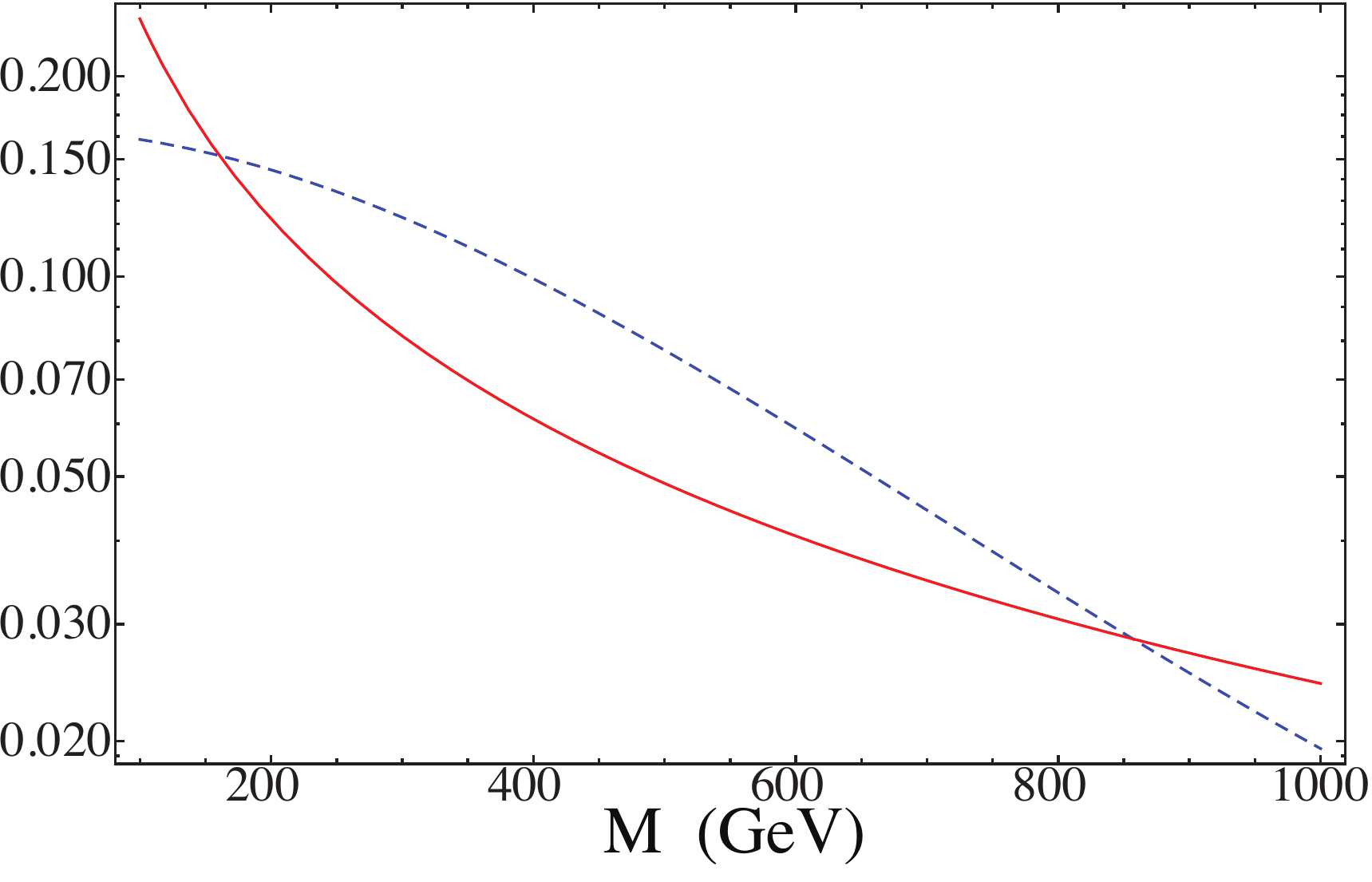}

\caption{Numeric solution to Eq.~(\ref{eq:singletdoubletmix}) for $\epsilon = 0.16$ (left) and $\epsilon = 0.2$ (right). The blue dotted line is the left-handed side of Eq.~(\ref{eq:singletdoubletmix}) and the red line is the right-handed side. \label{fig:singletdoublet}}
\end{figure}

With these parameters, the action of the sphaleron in the early Universe is to create $(1-\epsilon^2/2)^2$ particles of mass $M$, and $\epsilon^4$ particles of mass $\epsilon^2 M$. The heavy states will then decay, leaving only the light dark matter in the Universe today. From Eq.~(\ref{eq:numberdensity}), the correct amount of dark matter is found when
\begin{equation}
\epsilon^2 f(\epsilon^2 M/T_D) + (1-\epsilon^2) f(M/T_D) = f(0) \frac{\rho_{\rm DM}}{\rho_{\rm B}} \frac{m_{\rm proton}}{\epsilon^2 M}. \label{eq:singletdoubletmix}
\end{equation}
Solutions to this equation only exist when $\epsilon \gtrsim 0.16$, in which case $M \sim 400\gev$, resulting in dark matter with mass $\epsilon^2 M \sim 9 \gev$ (see Fig.~\ref{fig:singletdoublet}). In such a mass regime, the strictest constraints come from measurements of the invisible $Z$ width, which require $\epsilon \leq 0.23$. Amusingly, with the choice of $\epsilon = 0.2$, two solutions to Eq.~(\ref{eq:singletdoubletmix}) exist, $M\sim 150\gev$ and $M\sim 850\gev$ -- natural masses, in light of our expectations of new physics at the weak scale. The first solution has a dark matter candidate with mass of $6\gev$ and the appropriate cross section to explain the CoGeNT and DAMA/Libra direct detection anomalies. However, adding non-negligible mass terms $m_{12}$ and $m_{13}$ can alter the mass relations of the light state, and we have not proposed any mechanism that explains our arbitrary choice of parameters.

%While electro-weak precision tests (EWPT) are potentially a constraint,
%independent of mass i thought?? strumia claims  there's a 1/m^2 dependence
%the contribution is negligibly small \cite{Cirelli:2005uq}. We also comment on detectability. Collider signatures are difficult due to the large mass. The charged state could possibly be visible as a short `stub' track in the inner tracker before it decays into a low energy pion and the invisible neutral state \cite{Cirelli:2005uq,Buckley:2009kv}. However, the production rate for $2\tev$ weakly charged fermions makes this a difficult proposition.

\subsection{$\mathbf{SU(2)_R}$ Sphalerons}

We have seen that $SU(2)_L$ sphalerons are a promising way to transfer an initial $X$ asymmetry into baryons. Such processes are already known to exchange baryons and leptons, so no new interactions are required. However, additional singlet fields are necessary  to avoid direct detection bounds.   In addition, the non-relativistic solution to the chemical equilibrium forces the dark matter mass to be very large compared to the sphaleron decoupling temperature $T_D$, requiring a large Yukawa coupling at the border of perturbativity or the presence of additional right handed states.   Transfer via a sphaleron of some additional gauge group avoids the first limitation, but does not solve the second. 

A new non-abelian gauge group with chiral representations in both the SM and the dark sector would move an $X$ asymmetry into the visible sector, analogous to the $SU(2)_L$ case.  As in the $SU(2)_L$ example, the right-handed sphaleron would be active down to a temperature $T_D$ which we assume to similar to the vev of the $SU(2)_R$-breaking Higgs field $T_D \sim \langle \Phi\rangle \equiv v_R$. Experimental constraints tell us that $M_{W_R} > 4\tev$, assuming that the gauge bosons couple to all three generations of the SM. As these masses are related to the vev by $v_R = 2M_R/g_R$, assuming a perturbative gauge coupling, $v_R$ (and thus $T_D$) must be heavier than $\sim 8\tev$. In a SUSY model, cancellations in the FCNCs can reduce the bound to $4\tev$ \cite{Zhang:2007qma}. If only the third generation coupled to the new gauge force, then the bounds are much weaker. Assuming coupling to $\gamma/Z$, LEP-II places a bound of $105\gev$ on $M_{W_R}$, and thus $T_D$ must be greater than $\sim 200\gev$, similar to the $SU(2)_L$ example.

Repeating the calculation of chemical equilibrium, the correct dark matter abundance is found when
\begin{equation}
f(m_X/T_D) =  \frac{3f(0)}{N_f'N_X^2} \frac{\rho_{\rm DM}}{\rho_B} \frac{m_{\rm proton}}{m_X}. \label{eq:su2rsolution}
\end{equation}
where $N_f'$ is the number of SM generations coupling to $SU(2)_R$. For $N_f = 3$ and $N_X=1$, $T_D \sim 8\tev$ and $m_X \sim 52\tev$. For only the third generation coupling, the non-relativistic solution for $N_X =1$ is $m_X \sim 1500\gev$. In both cases, the Yukawa couplings $y_X$ are again ${\cal O}(10)$.

\subsection{Left-Right Annihilation}

In the previous sections we found that minimal sphaleron models require a non-perturbative Yukawa coupling for the dark matter candidate. In the $SU(2)_L$ case, we are able to avoid this constraint and evade direct detection bounds by the addition of light singlets. Here we present a second model that avoids the large Yukawa problem; one that establishes the right dark matter density  even if $m_X/T_D$ is much less than ${\cal O}(10)$. Asymmetries are still exchanged via $SU(2)_R$ sphalerons, but we assume in addition to a single left-handed doublet charged under $SU(2)_R$ there is also  a second right-handed doublet as well as a massive singlet.\footnote{Graham Kribbs \cite{graham} has considered similar ideas.} In the absence of the singlet,  the sphaleron would create a left-handed particle $X_L$ and destroy a right-handed particle $X_R$. As a result, the $X$ number would not be changed by the sphaleron, and no Xogenesis could occur.

With the addition of the singlet, the three fields can mix. For simplicity, we assume only singlet-$X_R$ coupling, so the neutral Lagrangian is
\begin{equation}
{\cal L} \supseteq m_{LR} X_L \bar{X}_R + y X_S \phi \bar{X}_R + m_S X_S \bar{X}_S +\mbox{h.c.} \label{eq:leftrightlag}
\end{equation}
Including a Higgs vev, the mixing mass term becomes $m_R \equiv y v$. Assuming $m_{LR}, m_R \ll m_S$, the two light eigenstates ($X_1$ and $X_2$) are $m_{LR} \pm m_R^2/2m_S$ and the heavy state $X_3$ is $m_S + m_R^2/m_S$.

The primarily singlet particle we define as $X_3$ has a small mixing angle $\sin\theta \approx m_R/m_S$ with the right-handed doublet. Thus, if $m_S \gg T_{\rm sphaleron}$, when $T \gtrsim T_{\rm sphaleron}$, $X_3$ freezes out and the sphaleron creates one unit of left-handed field (in a combination of $X_1$ and $X_2$) and destroys $(1- \sin^2\theta)$ unit of right-handed doublet (again in a linear combination of $X_1$ and $X_2$). Thus, after the freeze-out of the heavy state, the sphaleron changes $X$ number by $\sin^2\theta = m_R^2/m_S^2$. 

The linear combination $B-N_f\sin^{-2}\theta X$ is preserved by the sphaleron, and so
\begin{equation}
\mu_X = -\sin^2\theta \mu_{u_L}. \label{eq:lrmu}
\end{equation}
Again, assuming $m_S \gg T_{\rm sphaleron} \approx v_R$, which allows us to use $m_X \sim m_{LR}$ we can relate the number density of $X$ to that of baryons once the sphaleron decouples:
\begin{eqnarray}
n_X & = &  \frac{\sin^2\theta}{3}\frac{f(m_{LR}/T_{\rm sphaleron})}{f(0)} n_B \\
f(m_X/T_{\rm sphaleron}) & = & \frac{m_S^2}{2m_R^2} \frac{\rho_{\rm DM}}{\rho_{B}} \frac{m_{\rm proton}}{m_X} \label{eq:lrconstraint}.
\end{eqnarray}

\begin{figure}[ht]
\includegraphics[width=0.4\columnwidth]{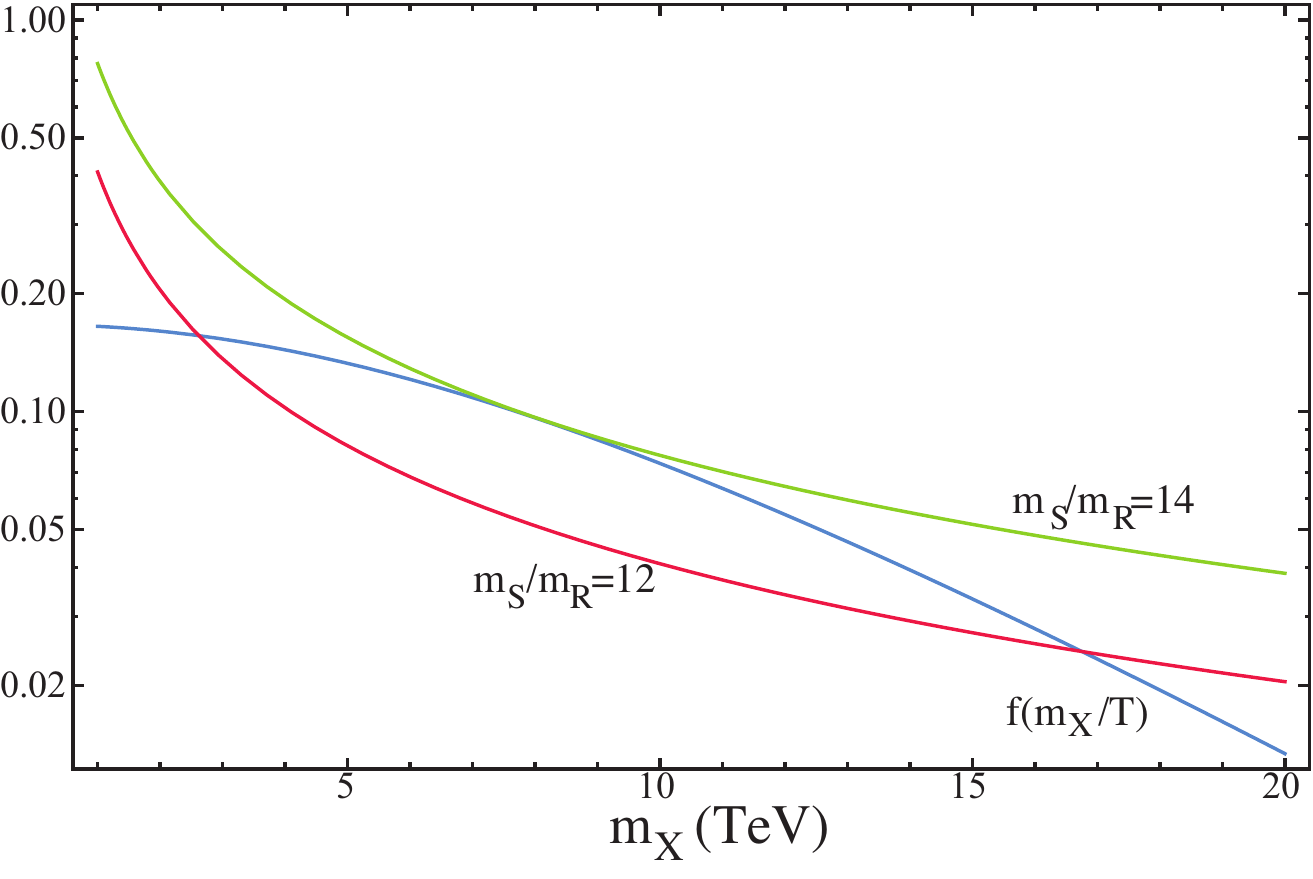}\hspace{2em}\includegraphics[width=0.4\columnwidth]{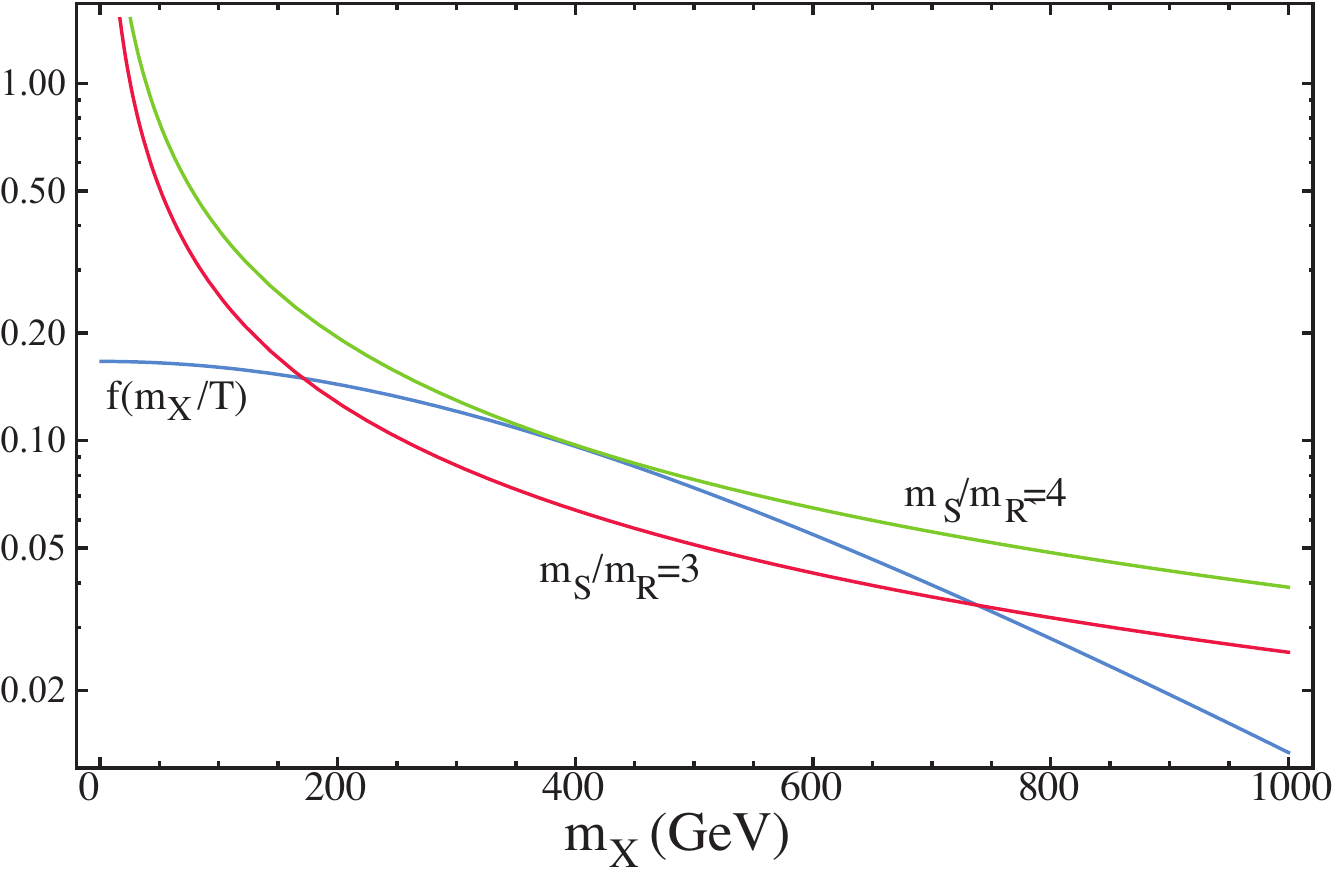}

\caption{The left- and right-handed sides of Eq.~(\ref{eq:lrconstraint}) showing the numeric solutions dark matter mass $m_X$ assuming two values of $m_S/m_R$ for $T_{\rm sphaleron} = 4\tev$ (left figure) and $200\gev$ (right figure). Shown are the critical solutions, when $m_S/m_R = 14(4)$ ($T_{\rm sphaleron} = 4\tev(200\gev)$) and an example of solutions with a smaller value of $m_S/m_R$. \label{fig:leftright}}
\end{figure}

Since the sphaleron creates only a small change in $X$ number (relative to baryon number), the thermal suppression of the dark matter number density need not be very large. This allows for lighter dark matter than the previous case. If all three generations are coupled to $SU(2)_R$, then the lowest $v_R$ can be is $\sim 4 \tev$. In this case, solving Eq.~(\ref{eq:lrconstraint}) numerically, we find solutions exist when $m_S/m_R < 14$. At the critical value of the ratio, only one solution for the dark matter mass exists, $m_X \sim 8\tev$. For smaller ratios, two solutions exist, one lighter and one heavier than $8\tev$ (see Fig.~\ref{fig:leftright}).

If we assume that only the third generation couples to $SU(2)_R$, and so $T_{\rm sphaleron} \sim v_R$ can be as low as $200\gev$. Again solving for $m_X$ in Eq.~(\ref{eq:lrconstraint}), we find that no solution exists if the ratio $m_S/m_R \gtrsim 4$. At this critical point, $m_X \sim 400 \gev$. As before, with smaller ratios two solutions exist. 

This set of models avoids the problem of large Yukawa couplings, as the heavy mass $m_S$ does not arise from the $SU(2)_R$ Higgs vev. Were it not for the direct detection bounds, a similar method could apply to Xogenesis models involving a $SU(2)_L$ sphaleron transfer: however the singlet component of $X_1$ necessary  to achieve the correct dark matter density is small, and so would not suppress the direct detection cross section by the large factors needed to evade the current limits.

\subsection{${\mathbf B}$ Violating Operators}

We next consider interactions that can transfer dark matter to baryon number directly -- that is, operators that explicitly violate $X$ number as well as either $B$ or $L$. In each case, for explicitness, we consider minimal operators in a SUSY scenario as presented in Ref.~\cite{Kaplan:2009ag}, though we expect many other possibilities exist. For baryon violating operators,  we assume a superpotential of the form
\begin{equation}
W \supseteq \frac{1}{M^2} X X u d d. \label{eq:bviolW}
\end{equation}
In the early universe when temperatures are above $T_D$, this interaction allows the $X$ asymmetry to be transferred into $B$ via squarks, which decay quickly into quarks and neutralinos. Assuming a fermionic $X$ and $m_{\tilde{q}} > m_X$, the cross section for this process is approximately
\begin{equation}
\langle \sigma v \rangle \sim \frac{\pi m_X^2}{(16\pi)^2 M^4}e^{-(3m_{\tilde{q}}-2m_X)/T}  \label{eq:bviolcross}
\end{equation}
where the exponential suppression comes from requiring two $X$ particles to be far enough out in their velocity distribution to have sufficient  energy to create three squarks.
The rate for the resulting $B-X$ transfer is
\begin{equation}
\Gamma = n \langle \sigma v \rangle \sim \left(\frac{m_XT}{2\pi}\right)^{3/2} e^{-m_X/T}\frac{\pi m_X^2}{(16\pi^2)^2M^4} e^{-(3m_{\tilde{q}}-2m_X)/T} = \left(\frac{m_XT}{2\pi}\right)^{3/2} \frac{\pi m_X^2}{(16\pi^2)^2M^4} e^{-(3m_{\tilde{q}}-m_X)/T}. \label{eq:bviolgamma}
\end{equation}
This process decouples at some temperature $T_D$, when $\Gamma$ is equivalent to the expansion rate of the Universe, $\sim 20 T^2/m_{\rm pl}$. We focus on the more experimentally accessible examples with low UV completion scale $M$. These models yield relatively low values of $T_D$ as well \cite{Kaplan:2009ag}.

As with the generic sphaleron scenario outlined previously, the chemical equilibrium equations for the baryon violating operator force $T_D \sim m_X/10$. Therefore, the free parameters in Eq.~(\ref{eq:bviolgamma}) are $m_X/M$, and $\Delta m \equiv 3 m_{\tilde{q}} - m_X$. The solution for $T_D$ depends linearly on $\Delta m$ but only logarithmically on $m_X/M$; assuming that $M$ is not many of orders of magnitude larger than $m_X$ we find a solution when $\Delta m \sim 45 T_D$. This implies that $m_{\tilde{q}} \sim 20 T_D$. 

This still has not set an overall mass scale. From the lower bound on squark masses of $\sim 400\gev$ \cite{pdg},  $T_D\gtrsim20\gev$ and $m_X \gtrsim 200\gev$.The scale $M$ in these scenarios was assumed to be $1\tev$. Smaller values are inconsistent with the assumption that Eq.~(\ref{eq:bviolW}) is an effective operator. Larger values are possible and only logarithmically affect $\Delta m$ and $T_D$.

If the dark matter is instead a scalar, then the superpotential Eq.~(\ref{eq:bviolW}) will allow interactions between two scalar $X$, a squark and two quarks. The calculation for the freeze-out temperature will proceed in a similar manner to the fermionic case, with the replacement of $3m_{\tilde{q}} - 2m_X$ with $m_{\tilde{q}}-2m_X$. For a given squark mass, this allows for slightly lower $T_D$ and $m_X$; for example, a $400$~GeV squark has $T_D \sim 6\gev$ and $m_X\sim 60\gev$ with only a logarithmic dependence on $M$.

The symmetric component of dark matter must of course be eliminated. The $X-B$ transfer term in the superpotential Eq.~(\ref{eq:bviolW}) by itself does not provide sufficient annihilation in the early Universe, so some additional structure must be present. The two most obvious possibilities are annihilation via (pseudo-)scalars or through vector bosons. We consider them in turn; both are capable of removing the symmetric component, though the vector scenario may require additional structure.

We first consider the annihilation of the symmetric dark matter through the lightest pseudo-scalar $a$ in the NMSSM, as discussed in Ref.~\cite{Kaplan:2009ag}. We require the thermal abundance of dark matter to be much less than the full $\Omega_{\rm DM}h^2 \sim 0.1$ observed today. This can be roughly translated as the requirement that the annihilation cross section in the early Universe must have been $\gg 1$~pb. As demonstrated in Ref.~\cite{Kaplan:2009ag}, annihilation mediated by $a$ has a sufficiently large cross section provided the pseudoscalar vev is $\ll 200\gev$, which may lead to interesting supersymmetric physics. This interaction also allows the possibility of a direct detection signal, with a $X$-nucleon elastic cross section of
\begin{equation}
\sigma(X n \to X n) = 6.0 \times 10^{-43} ~\mbox{cm}^2 \times g_{XXh}^2 \left( \frac{m_h}{100 \gev}\right)^{-4} \label{eq:higgsdirectdetection}
\end{equation}
which, for perturbative Higgs-$X$ couplings of ${\cal O}(0.1)$, gives a direct detection cross section about an order of magnitude below the current best bounds from XENON100 \cite{Aprile:2010um} and CDMS-II \cite{Ahmed:2009zw}.

The second possible annihilation scenario is via a new broken `dark' gauge group, either abelian or non-abelian (the dark forces cannot be massless \cite{Buckley:2009in}). In either case, the dark matter, annihilating through a $Z'$ or $W'$, must either go to lighter dark states (which may include the $W'$ itself) or, through some small mixing, to SM fields. There are significant constraints on the latter scenario, while the former requires some new symmetry to prevent the dark matter from decaying directly to the lighter state in the dark sector.

Assuming a low  ($\lesssim \gev$) scale mass for the gauge bosons, the thermal relic abundance is much smaller than the total dark matter density when \cite{Ackerman:2008gi,Feng:2010zp}
\begin{equation}
\alpha' \gg 10^{-3} \left(\frac{m_X}{100\gev} \right). \label{eq:aprimelimit}
\end{equation}
If the annihilation is into some new dark state (or in the case of non-abelian gauge groups, $XX \to W'^* \to W' W'$), then there need be no direct contact with the SM fields at low energies, and so the possibility of direct detection are significantly reduced. However, if the gauge group is abelian, then generically we expect some small kinetic mixing between the $Z'$ and the photon parameterized by $\epsilon \sim 10^{-2} -10^{-6}$ \cite{Bjorken:2009mm}. In such a scenario, the requirement of Eq.~(\ref{eq:aprimelimit}) still applies, with $\alpha' \to \epsilon \sqrt{\alpha \alpha'}$.

If the $Z'$ mixes with the visible sector, then the dark sector can be probed not only by direct detection, but also by beam experiments \cite{Bjorken:2009mm,Essig:2010xa}. The current limits from direct detection are very constraining. For example, if dark matter consists of Dirac fermions, then the combination $\sqrt{4\pi \alpha' \epsilon}$ must be less than $\sim 10^{-13}$ \cite{Batell:2009vb}. While this is in tension with the requirements for efficient annihilation, it does not necessarily rule out $Z'$-mediated annihilation.

Note that, in principle, the dark matter could scatter in direct detection experiments via the same baryon violating operator that  generates the asymmetry. While this is an attractive possibility that does not require the addition of new elements to the theory,   the operator in question is dimension-12 (four fermion to four fermion scattering) and not visible, even with the low scale of $M$. 

We note an interesting possible signature of baryon violating models first pointed out by Ref.~\cite{Davoudiasl:2010am}: annihilation of baryons with dark matter. While the interactions of Eq.~(\ref{eq:bviolW}) do allow for such diagrams, the large mass of $X$ compared to that of a nucleon, combined with the low velocities of dark matter in the halo means that such events would be kinematically suppressed in direct detection experiments (or neutrino detectors such as Super-K). 

As this baryon violating model is implemented in a SUSY context, we also have to verify that the neutralino energy density is reduced to a sufficiently low level. The $X-B$ violating interaction creates squarks, which quickly decay into quarks and neutralinos. This would result in a ratio of baryons to neutralinos of $\sim 3$, if no additional annihilation occurs. The neutralinos themselves will be in thermal equilibrium until their freeze-out temperature, typically $\sim m_{\tilde{\chi}}/20$. Assuming $100\gev$ masses, the neutralinos will still be in equilibrium at $T_D$, when the baryon violating operator decouples. As a result, we can use the standard calculations for neutralino WIMP relic abundance. The elimination of neutralinos favors a large wino component. From Ref.~\cite{Baer:2002ps}, for example, we see that  if the neutralino is primarily wino, the contribution to dark matter from $\tilde{\chi}$ is subdominant.  The large wino component of the neutralino seems a robust prediction of Xogenesis models based on low scale supersymmetry.

\subsection{${\mathbf L}$ Violating Operators}

Our final example of a $X$ transfer mechanism are lepton violating operators. Again, as in Ref.~\cite{Kaplan:2009ag}, we embed the lepton violation in a SUSY model, with the follow addition to the superpotential
\begin{equation}
W \supseteq \frac{1}{M} X X L H_u \label{eq:LviolW}
\end{equation}
This transfers the initial $X$ asymmetry into sneutrinos. After decaying to neutralinos and neutrinos, the asymmetry is then transferred into $B$ via the $SU(2)_L$ sphaleron. Assuming fermionic dark matter and $T_D > T_{\rm sphaleron} \sim 200\gev$, both $L$-violating operators and sphalerons are active throughout the $X-L$ transfer and so
\begin{equation}
\mu_X = \frac{33}{14} \mu_{u_L}. \label{eq:Lviolmu}
\end{equation}
This leads to the following equation which must be solved numerically
\begin{equation}
f(m_X/T_D) = \frac{14 f(0)}{11} \frac{\rho_{\rm DM}}{\rho_B} \frac{m_{\rm proton}}{m_X}. \label{eq:LviolF}
\end{equation}
Again, the the non-relativistic solution provides the correct amount of dark matter when $T_D \sim m_X/10$. As we have assumed $T_D > 200\gev$, we see that this scenario leads to very heavy dark matter, and consequently very heavy sneutrinos. 

As with the baryon violating operators, the decoupling temperature $T_D$ can be related to the masses by setting the rate of $XX \to \tilde{\nu}$ equal to the Hubble expansion. Again assuming an exponential suppression of the rate due to mass difference between two dark matter particles and the sneutrino $\Delta m \equiv m_{\tilde{\nu}}-2m_X$, the decoupling occurs when
\begin{equation}
\frac{1}{16\pi} \left( \frac{m_XT_D}{2\pi}\right)^{3/2} \left(\frac{v_u}{M}\right)^2 m_{\tilde{\nu}} \left( 1-\frac{4m_X^2}{m_{\tilde{\nu}}^2}\right)^{3/2} e^{-\Delta m/T_D} \sim \frac{20 T_D^2}{m_{\rm pl}}. \label{eq:LviolTD}
\end{equation}
With the vev $v_u \sim 200\gev$, $M\sim 10\tev$, and the previously stated assumptions on $T_D$ and $m_X$, we find that the sneutrinos themselves would need to be $\sim 30T_D$, that is, around $6\tev$.

If the dark matter is a scalar, then the $X-L$ transfer occurs through $XX \to \tilde{h}\nu$ scattering. The rate is again controlled by an exponential of $e^{-(m_{\tilde{h}}-m_X)/T_D}$; as in the fermionic scenario, the decoupling temperature depends only logarithmically on $M$. As in the fermionic case, we again find that the supersymmetric scale must be high; with a higgsino mass of $\sim 6 \tev$.

\subsection{Bleeding $X$ into $L$}

The previous subsection assumed the lepton operators decouple before the weak interaction sphalerons, implying heavy dark matter and sneutrinos. We now consider the alternative -- and more desirable -- option where  $T_D < m_X < T_{\rm sphaleron}$. Rather than suppress $X$ relative to $B$ number thermally, in  this scenario the lepton number violating operators will still be in equilibrium below the temperature at which electroweak sphalerons have shut off, allowing $X$ number to bleed off into neutrinos. For this reason, the $X$ number can be naturally less than $B$ number, even without a larger thermal suppression factor.   

The lepton violating operator is interesting in this respect. As the operator only provides chemical equilibrium between $X$ and $L$, baryon number is created only when the $L$ number is transferred via the $SU(2)_L$ sphaleron into $B$.   That means that $X$ and $L$ number can continue to be violated, even when $B$ number has already been established.   Above $T_{\rm sphaleron}$, the $X$ asymmetry is transferred to $B$ via a combination of the $L$-violating operators and the sphaleron. As there is no thermal suppression due to large $m_X$, this sets the $X$ and $B$ numbers to be roughly equivalent. For example, in a MSSM scenario,
\begin{equation}
n_X(T_{\rm sphaleron}) = \frac{11}{28} n_B. \label{eq:nxsphaleron}
\end{equation}

\begin{figure}[th]
\includegraphics[width=0.6\columnwidth]{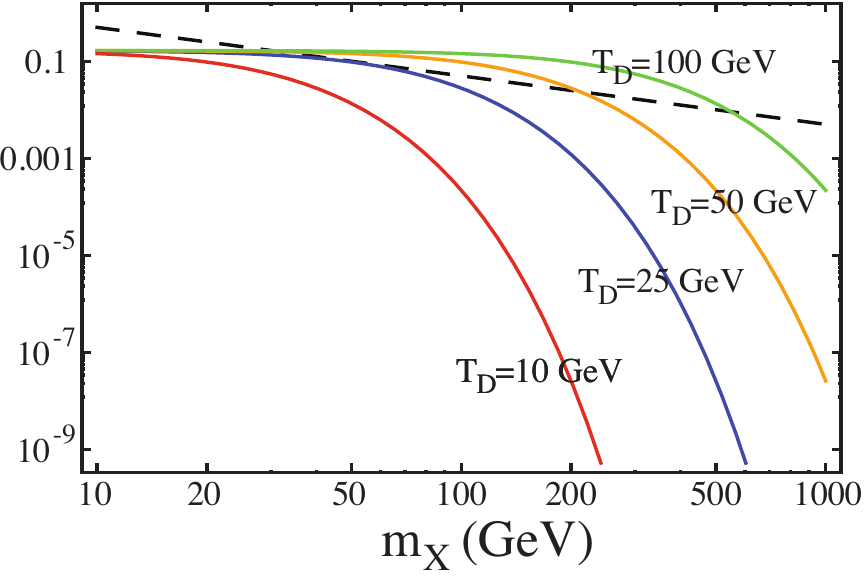}
\caption{Numeric solutions to Eq.~(\ref{eq:ftransfer}). The black dotted line is the right-hand side of Eq.~(\ref{eq:ftransfer}), while the left-hand side is shown assuming $T_D$ is $10$ (red), $25$ (blue), $50$ (orange), or $100\gev$ (green). For $T_D < 25\gev$, no solutions occur, while for larger values, two solutions exist.  \label{fig:transfer}}
\end{figure}

After this initial equilibration, the sphalerons shut off. However, the dark matter still can exchange particle number with $L$, via operators like that in Eq.~(\ref{eq:LviolW}). If $m_X \gg m_{\rm proton}$, then most of the dark matter will have to convert into leptons to avoid overclosing the Universe, that is the present number density of neutrinos must be $n_\nu \sim n_X(T_{\rm sphaleron})$. Again, using the MSSM scenario, this requirement translates into
\begin{equation}
f(m_X/T_D) \approx \frac{56}{66} \frac{\rho_{\rm DM}}{\rho_B} \frac{m_p}{m_X}. \label{eq:ftransfer}
\end{equation}
We solve numerically for different values of $T_D$. As seen in Fig.~\ref{fig:transfer}, for $T_D \lesssim 25\gev$, there are no solutions available. At $T_D \sim 25\gev$, there is a single solution at $m_X \sim 50\gev$. Above this, two solutions 
%?? why two?? do we have a scaling relation to explain ?? 
to Eq.~(\ref{eq:ftransfer}) are found. 

The first solution is at low mass, and corresponds to the relativistic solution to the chemical equilibrium equations, similar to the scenarios explored in Refs.\cite{An:2009vq,Cohen:2009fz,Cohen:2010kn,Kaplan:2009ag,Davoudiasl:2010am,Haba:2010bm,Shelton:2010ta}. Note our solution has a larger mass ($\sim 25$) than those found in the previous works ($\sim 5-10\gev$). This is easily understood, as bleed-off of number density into leptons forces each remaining dark matter particle to be heavier to make up for the loss. It is very interesting that this simple reversal of the hierarchy between $T_D$ and $T_{\rm sphaleron}$ can have such a large effect on the predictions of dark matter mass in a relativistic Xogenesis scenario.

The second solution is the non-relativistic one; requiring $m_X/T_D$ to be ${\cal O}(1)$. Notice that this is significantly lower than the ratio in the generic non-relativistic solutions we found in most other models, which are of ${\cal O}(10)$. As we have reduced the dark matter density by dumping additional particle number into the lepton sector, there does not need to be as much of an exponential suppression -- allowing a smaller $m_X/T_D$.

As in the baryon violating model, the neutralinos coming from the decay of the sneutrinos must be eliminated, requiring the $\chi$ to be primarily wino. In addition, as pointed out by Ref.~\cite{Kaplan:2009ag}, the UV completion of the Eq.~(\ref{eq:LviolW}) $L$-violating operator must contain new fields in either a singlet or doublet representation of $SU(2)_L$. If the latter, the symmetric component of dark matter can be removed via an intermediate doublet. This doublet can also induce a signal in  direct detection. The  annihilation cross section is
\begin{equation}
\langle \sigma v\rangle \approx \frac{1}{16\pi} \frac{y' m_X^2}{m_D^4}, \label{eq:doubletannihilation}
\end{equation}
here $y'$ is the doublet-lepton-$X$ coupling. To be efficient, $m_D/y'$ must be much smaller than $300\gev$, implying a very low scale $m_D$ ($\sim M$). The corresponding rate in direct detection experiments would be
\begin{equation}
\sigma(Xn \to Xn) \approx 10^{-46}~\mbox{cm}^2 \left(\frac{Z/A}{0.4}\right)^2 \left( \frac{m_X/y'}{100\gev}\right)^{-4}, \label{eq:doubletdirectdetection}
\end{equation}
assuming a ratio $m_D/m_X \sim 10$. This is approximately three orders of magnitude below the current CDMS-II/XENON100 bounds. Alternatively, the symmetric component of dark matter could be removed via either the singlet Higgs or dark gauge group methods discussed in the baryon violating mechanism. The former predicts direct detection rates approximately an order of magnitude below the current bounds, while predictions for the latter depend on the details of mixing between the dark gauge sector and the SM.

\section{Conclusion}
The near coincidence of scales that follows from dark matter relic thermal abundance (the ``WIMP Miracle''), has been the primary theoretical motivation in the field for many years. Though theoretically well-motivated, we should remember that it has not yet been experimentally proven. In light of the many recent anomalies from direct and indirect detection, which are difficult to reconcile with the expectations of WIMP dark matter, we should continue to seek alternative motivating principles.

The near coincidence between baryon and dark matter energy densities leads to to concept of asymmetric dark matter, in which dark matter consists of a particle without a sizable relic density of the corresponding antiparticle. To explain the coincidence of energy densities,, the relic number density of dark matter is related to that of baryons, which requires operators that violate baryon and dark matter number.  In this paper, we show that the energy densities can be appropriately related in a natural framework involving a weak scale dark matter candidate, in addition to the light asymmetric dark matter often considered (see {\it e.g.}~\cite{Cohen:2009fz,Shelton:2010ta,Davoudiasl:2010am,Haba:2010bm}). 

With such heavy dark matter, searches are difficult. Indirect detection requires additional structure to allow $X-\bar{X}$ oscillations in the late Universe, and only in specific cases will direct detection be expected. If the candidate is truly weak scale, it can be part of some larger sector detectable at the LHC, although identifying it as dark matter will be challenging. Nonetheless, given the uncertainty in the nature of dark matter, it is worth considering further this alternative as it seems to address one of the most striking features about the energy densities in the Universe.

\section*{Acknowledgements}
We would like to thank the Aspen Center for Physics for providing a wonderful opportunity for collaboration and discussion. We would also like to thank Kathryn Zurek, Graham Kribs, Patrick Fox, Dan Hooper, David Morrissey, Mark Wise, and Michael Ramsey-Musolf for helpful conversations.
We note that after our work began, several papers \cite{Davoudiasl:2010am,Haba:2010bm,graham,Shelton:2010ta} have presented examples with share some of the ideas we explore. MRB is supported by the Department of Energy, under grant DE-FG03-92-ER40701. LR is supported by NSF grant PHY-0556111.

\bibliography{xogenesis}
\bibliographystyle{apsrev}

\end{document}